\documentstyle[aps,pre,psfig]{revtex}

\begin{document}

\title{\noindent Journal de Physique III,7: (2) 303-310, (1997) 
\hfill  Section: Material Science \\ \ \\
Breakdown patterns in Branly's coheror}
\author{D. Vandembroucq$^{\dag,\ddag}$, A.C. Boccara$^{\dag}$ and
S. Roux$^{\ddag}$} 
\address{$\dag$ Laboratoire d'Optique, UPR CNRS 5,\\ $\ddag$
Laboratoire de Physique et M\'ecanique des Milieux H\'et\'erog\`enes, URA CNRS 857,\\ Ecole Sup\'erieure de Physique et de 
Chimie Industrielles,\\ 10 rue Vauquelin, 75231 Paris Cedex 05, France}
\maketitle

%\vspace{12pt}
%\centerline{(22 October 1996)}

\vspace{20pt}

\begin{abstract}
We use thermal imaging of Joule heating to see for the first time
electrical conducting paths created by the so-called Branly effect in a
two-dimensional metallic granular medium (aluminium). Multiple breakdowns are
shown to occur when the medium is submitted to high voltage increases
(more than 500 V) with rise times close to one hundred of microseconds. 
\end{abstract}

%\begin{multicols}{2}

\section{Introduction}
In 1890 Edouard Branly reported a surprising property of certain metal
powders~: when a radio source is approached to a generally
non-conducting assembly of iron or aluminum grains, the assembly
becomes conductive, and remains conductive even after the source is
removed \cite{Branly90}. A similar phenomenon may occur if the metal
powder is submitted to a DC voltage: as soon as the voltage exceeds a
certain threshold level, the assembly becomes conductive and remains
so when the voltage is removed or reduced \cite{Branly90,Onesti}. The
process is not however entirely irreversible; a small (mechanical)
shock is sufficient to turn the assembly back to its original,
insulating state. The radio-induced conductivity of this set-up (also
called Branly's coheror) was used in the first wireless telegraph
receivers.

Despite the recent development of physics of granular media, the
behavior of Branly's coheror is not completely elucidated. Several problems are indeed involved: the description of metallic
electrical contacts (surface roughness, oxide layer)~\cite{Boyer91}, the
distribution of local stress in a granular medium~\cite{Troadec91} and the
propagation of electromagnetic waves through a granular
medium in the case of the radio induced conductivity.

Although a complete physical understanding of the phenomenon is to be
found, some elements of information have been firmly established:
without any oxide layer at the grains' surface, this effect does not
exist; a graphite powder, a precious metal powder, an aluminium powder
previously reduced with hydrogen all directly
conduct~\cite{Kara75,Kama75,Kara90}; $1/f$ noise and giant fluctuations of
current~\cite{Kara75,Kama75,Kara90} are clear precursory to the transition to
the conductive state; the conduction seems to be due to filaments of
grains linked together (maybe by microfusions). This last point was
illustrated by an experiment at the Palais de la d\'ecouverte in Paris
\cite{Chesneau91}. The latter consisted of putting a point electrode
in a metallic cup filled with grains, the filings became conductive
when the voltage applied between the electrode and the cup was high
enough. Then lifting up slightly the electrode, it was possible to
pull up a chain of grains linked together.

In the following we present a visualization of the Branly effect
(section 2). This method is  used in section 3 to study the breakdown
patterns occurring in a two-dimensional metallic granular medium when
the latter is submitted to a high-voltage increase of varying rise
time. The results are then discussed in section 4.

\section{Thermal imaging: a visualization method of two-dimensional
  Branly's coheror} 

\subsection{Principle}

Up to now, no direct visualization of the phenomenon has been achieved.
Connections between grains in the assembly are very tenuous and thus
as one tries to extract the filament of connected grains, most
probably the structure will be broken.
Searching for links in an assembly of grains is not easy and not all links
have to be part of a conductive path. Moreover the mechanical cohesion
between grains is extremely small. Infrared imaging provides a solution
to this problem: if a conductive path exists, there are energy losses
in contacts by Joule heating and thus elevation of temperature. Even
if the metal generally poorly emits in the mid IR, the oxide layer  makes the grains emitting. When a current
flows through the grains, the local heating generates an infra-red image of the
conductive path(s) provided that  the assembly of grains remains
two-dimensional. We performed many experiments in order to study the
influence of different experimental parameters on the structure and the position of the paths.

\subsection{Realization}

\begin{figure} 
\centerline{\psfig{file=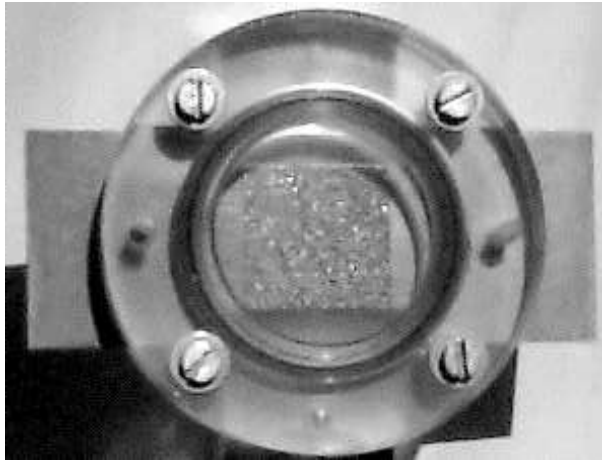,width=8.0cm,angle=-90}}
\caption{Experimental set-up}
\label{cellule}
\end{figure}

We have deposited one monolayer of aluminium grains (diameter 400-500$~\mu $%
m) in a $20$~mm square milled depression made in a thick disc of
plexiglass and bounded by two flat copper
electrodes. The layer of grains and the flat electrodes have been covered by
a 40~mm diameter circular sapphire window (which transmits both visible and
mid IR light below $5 \mu m$). As it can be seen on figure
\ref{cellule}, the set-up is maintained with a rubber O-ring screwed to the lower
part.

The conductive state is created with a high voltage generator (up to 1500~V)
which can not deliver current through low resistances such as those
obtained in the medium in its conductive state. We use a low voltage
generator for the visualization. The latter is done with an IR camera
$128\times128$  (AMBER 4128, InSb detectors with a cut-off at 6.5
$\mu$m) which allows us in the configuration used to reach a 
resolution of 200~$\mu$m by pixel. Between two consecutive visualizations,
we apply mechanical shocks to the set-up in order to break the
links previously created between the grains. Although not
quantitative, the latter process was checked not to induce
correlations in breakdown patterns, what suggests that most
connections between grains are broken.

\section{Simple and multiple breakdowns}

\begin{figure}[t]
\centerline{\psfig{file=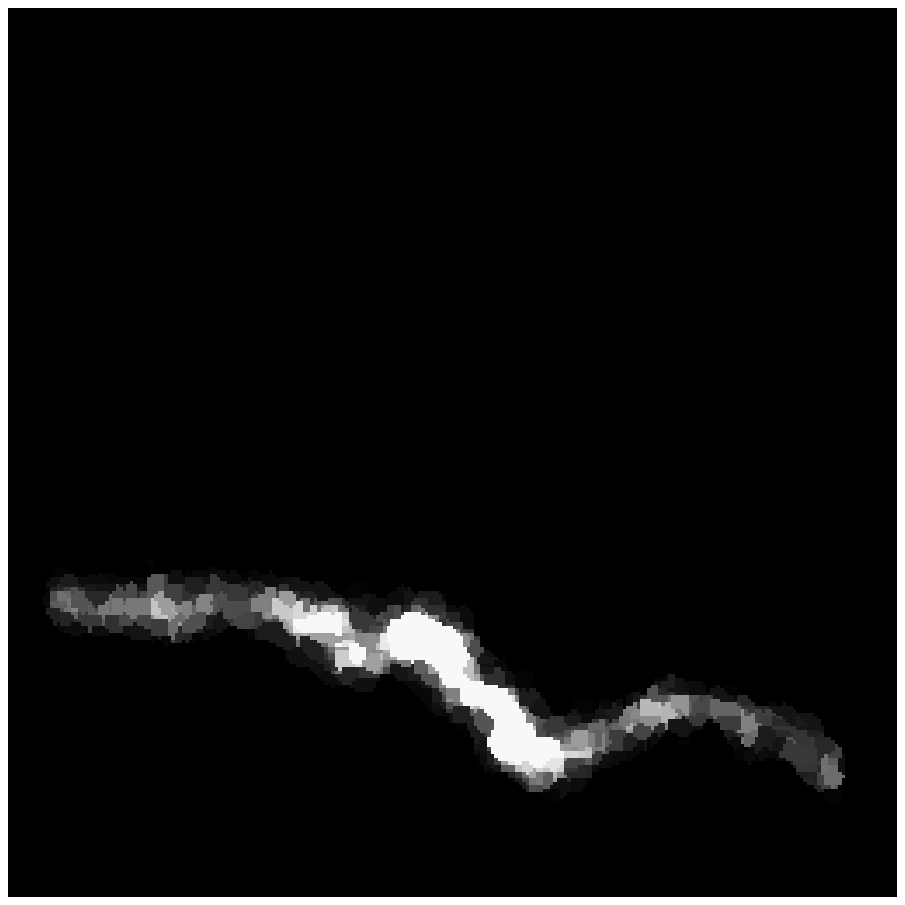,height=4.5cm,width=4.5cm} \hfill \psfig{file=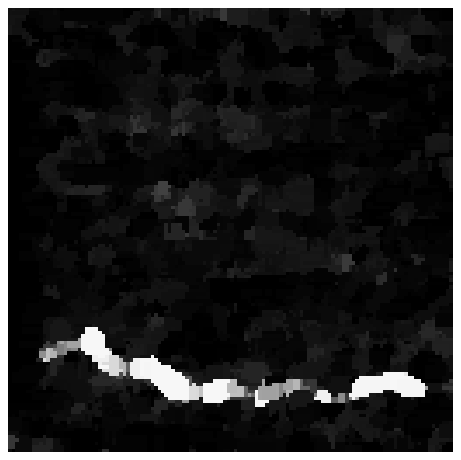,height=4.5cm,width=4.5cm} \hfill \psfig{file=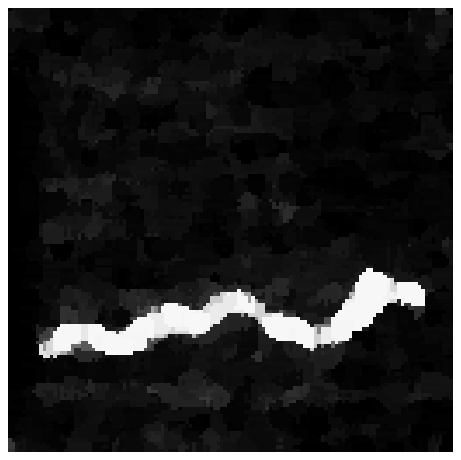,height=4.5cm,width=4.5cm}}
\vspace{0.25cm}
 \hfill 
\centerline{\psfig{file=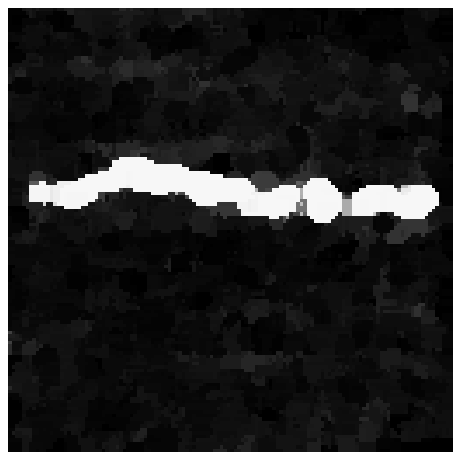,height=4.5cm,width=4.5cm}  \hfill \psfig{file=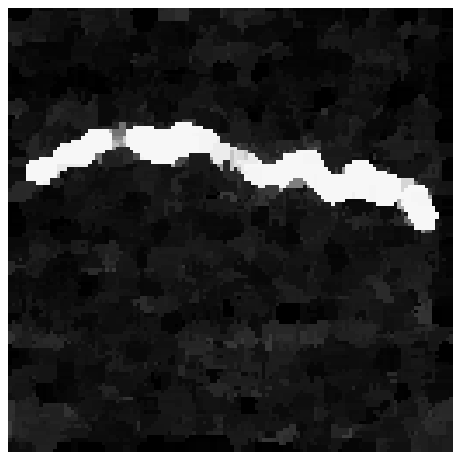,height=4.5cm,width=4.5cm} \hfill \psfig{file=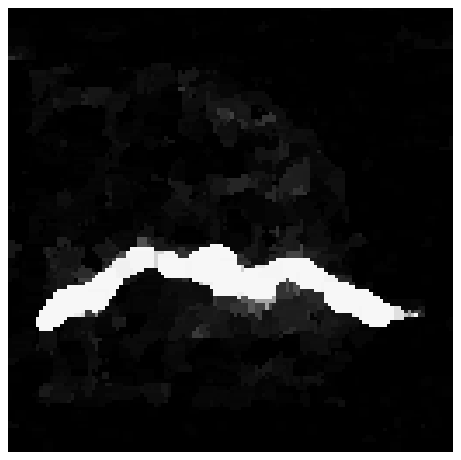,height=4.5cm,width=4.5cm}}
\vspace{0.25cm}
\caption{electrical conduction paths in a two-dimensional system of
  aluminium grains. The visualization is achieved with an infrared
  camera. The background has been subtracted so that the vertical
  black bands on left and right sides correspond to the copper
  electrodes.}
\label{single}
\end{figure}

The system of about 2000 aluminium grains we have used is sufficient
to recover the main features described by Branly at the end of the
last century. Although the two copper electrodes are only separated by
a few tens of grains, an electrical impedance measure of the system in its
original state gave us  $R=25$ M$\Omega$
and $C=0.5$ pF. After imposing a DC voltage of about 
500 V, this resistance
generally decreases down to about 100 $\Omega$. When a current of a few
mA flows through the coheror in this conducting state it becomes
possible to see a conducting path (see figure \ref{single}). These
images have been obtained by subtracting the background ({\it i.e.}
the image of the same system when no current flows through). The
orientation of the cell is the same as in figure \ref{cellule}. The
two vertical black bands on the left and right sides correspond thus
to the electrodes. The conducting path appears as a bright curve
connecting both electrodes. One can clearly see intensity contrasts
inside each path. As the current is imposed, these intensity contrasts
are due to the resistance contrasts that exist between the different
bonds linking the grains of the path. The brightest spots correspond
thus to the most brittle bonds.

Beyond the simple visualization, this experiment allowed us to confirm
the great dependence of the Branly effect on the stress distribution
in the granular medium. The latter is indeed known to be very
inhomogeneous \cite{Travers87}, which should affect the quality of the
electrical contacts. The specific case of spheroidal particles covered
by a soft shell - typically an oxide layer - while the inside is rigid
was recently studied by de Gennes \cite{deGennes96} and could be very
well adapted to the study of the Branly effect.

\begin{figure}[b] 
\centerline{\psfig{file=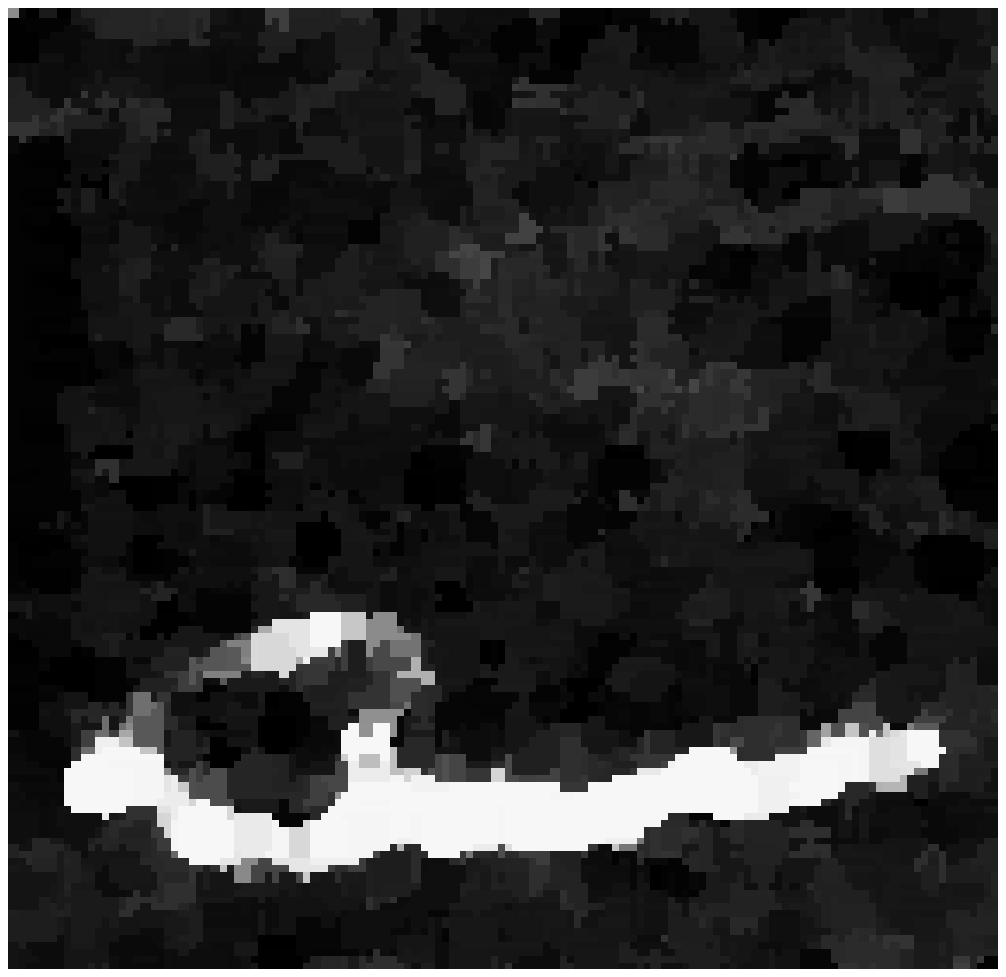,height=4.5cm,width=4.5cm} \hfill \psfig{file=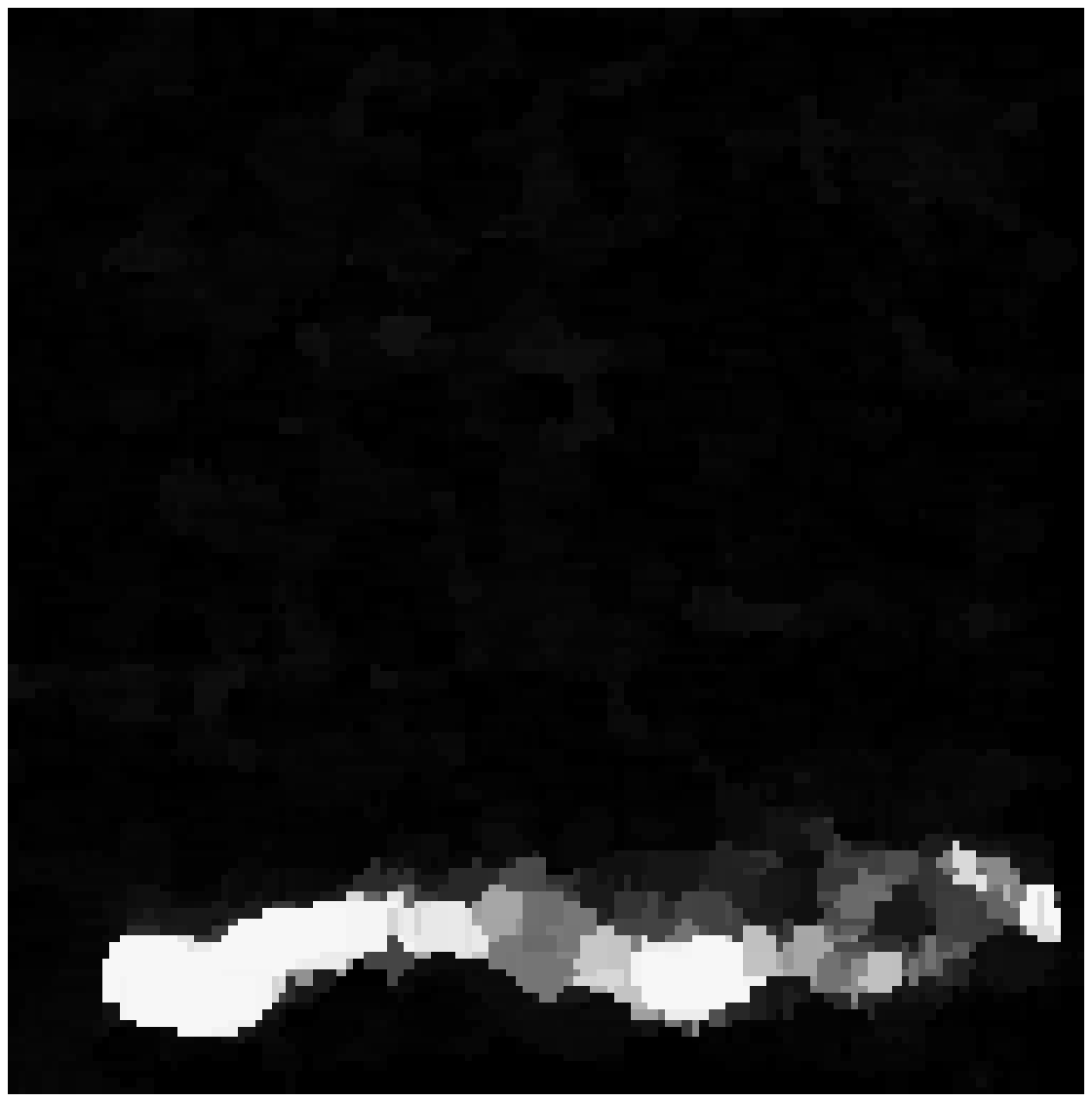,height=4.5cm,width=4.5cm} \hfill \psfig{file=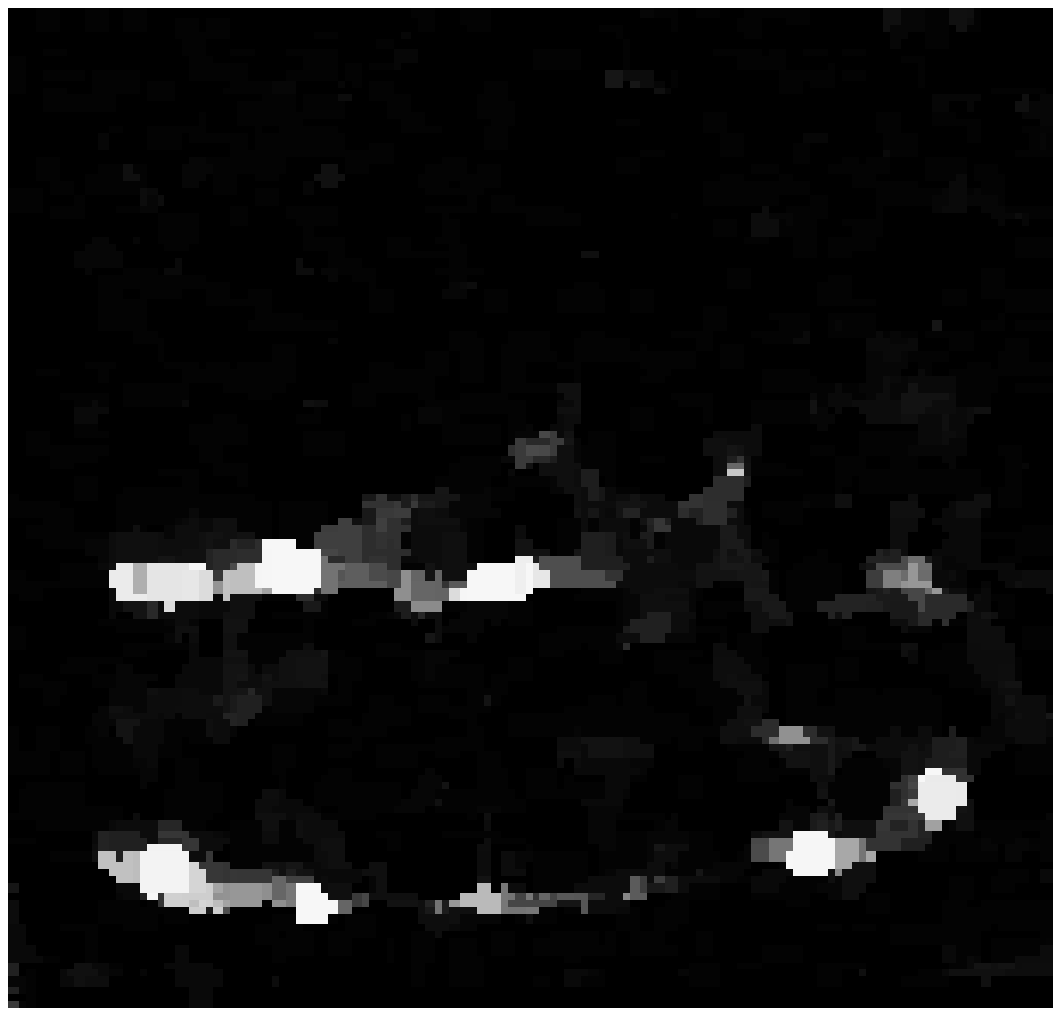,height=4.5cm,width=4.5cm}}
\vspace{0.25cm}
 \hfill 
\centerline{\psfig{file=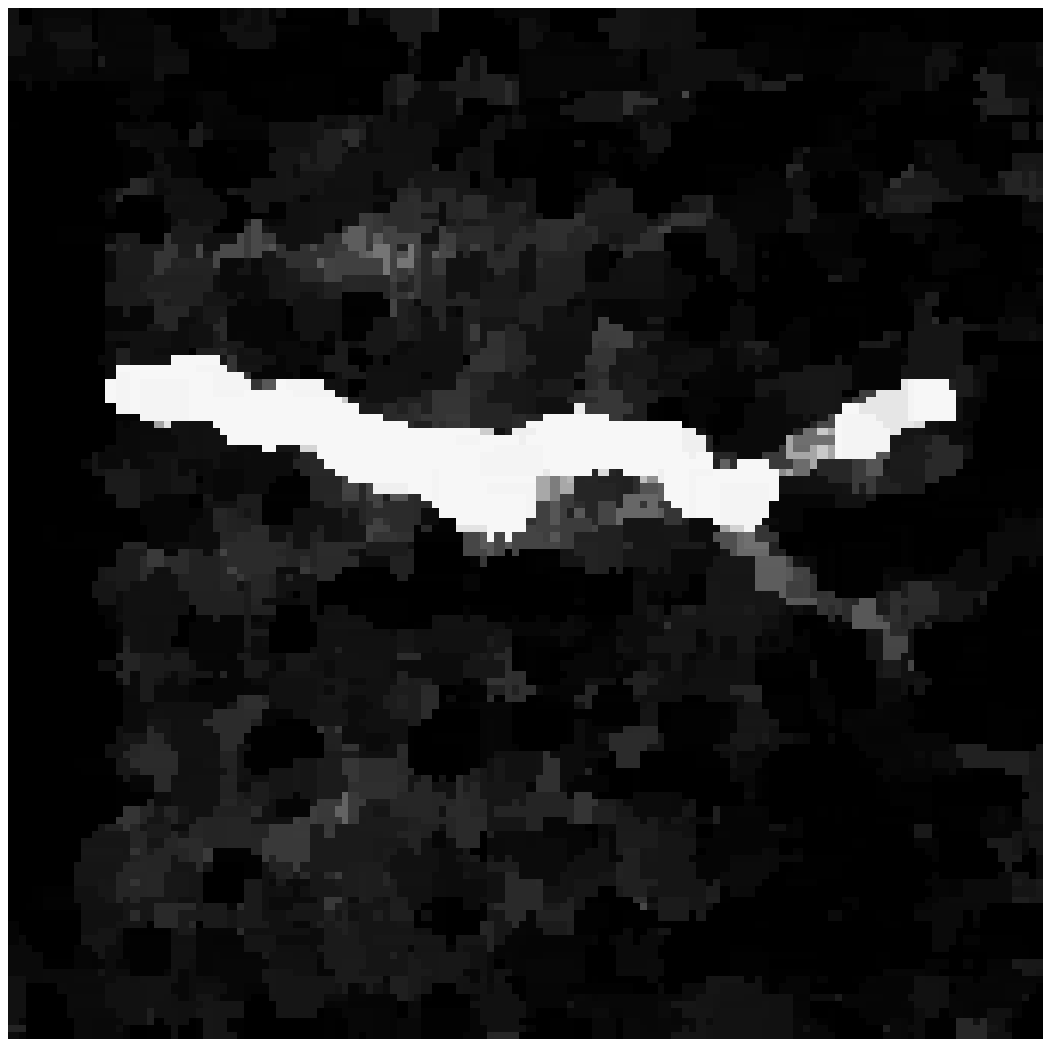,height=4.5cm,width=4.5cm}  \hfill \psfig{file=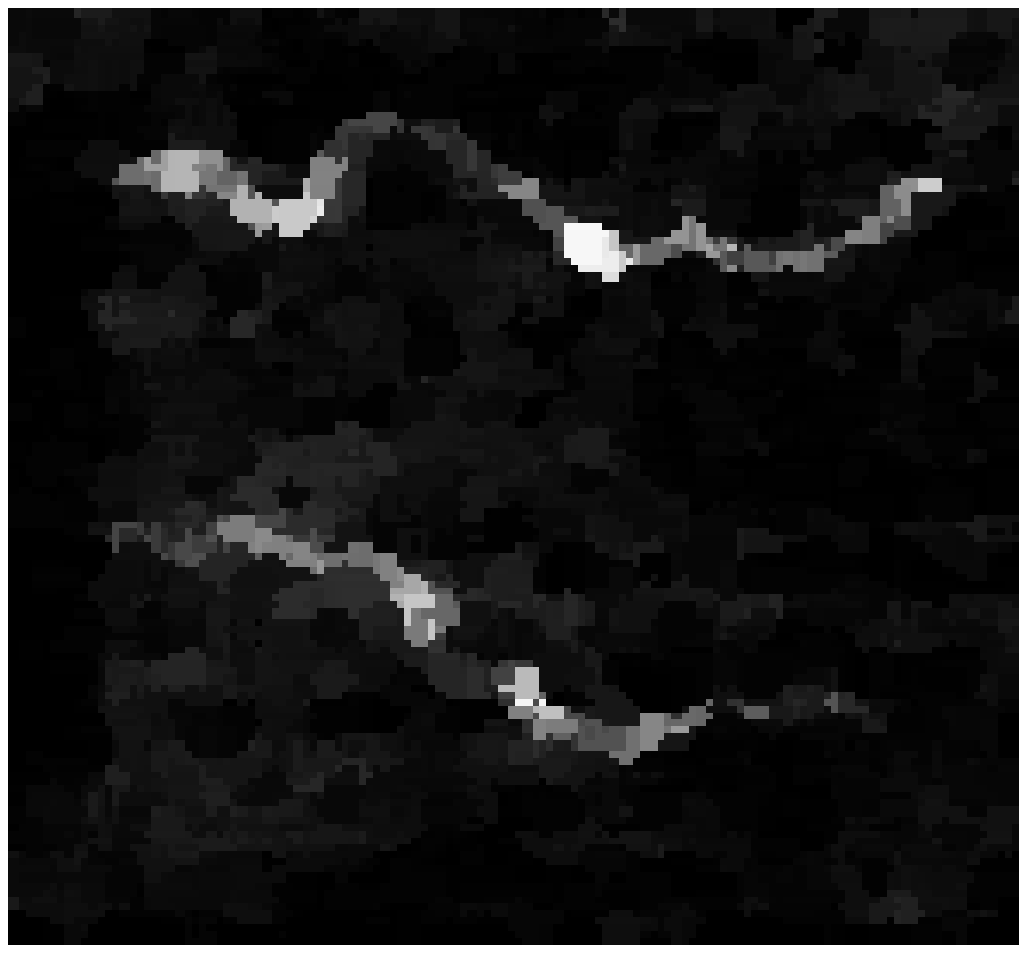,height=4.5cm,width=4.5cm} \hfill \psfig{file=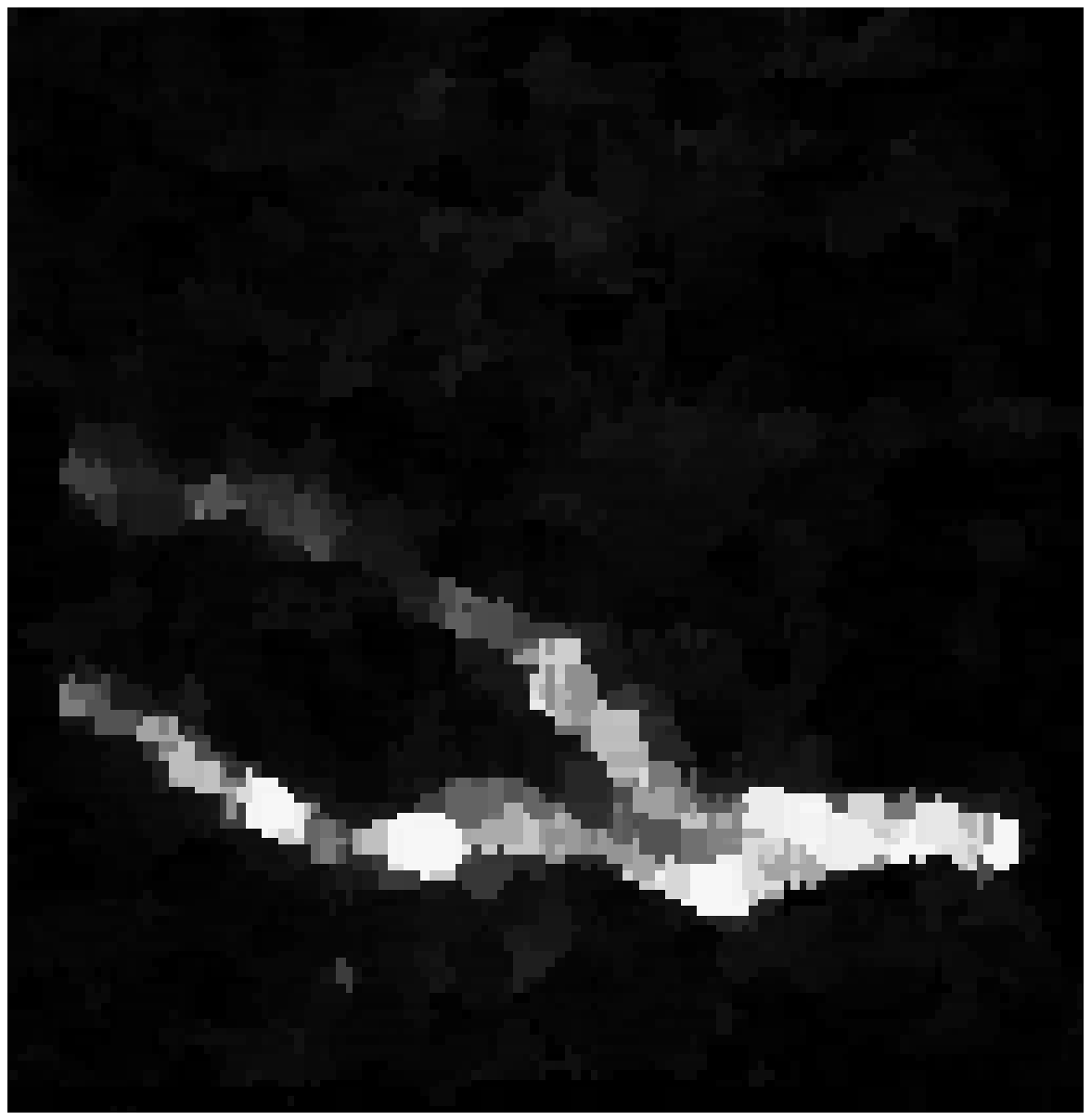,height=4.5cm,width=4.5cm}}
\vspace{0.25cm}
\caption{``Branly paths'' created with a maximum voltage of  500 V and
  a rise time of about 100~$\mu$s.}
\label{mult}
\end{figure}

  Varying the position
of the layer of aluminium grains, we have thus noted that in case of
vertical position, the conducting path was almost always created in
the lower part while it was spatially uniformly distributed in case of
horizontal position. The very low stress induced by the weight of the
upper grains suffices to change dramatically the electrical contacts
distribution in the medium and the creation of the conducting path.

\begin{figure} 
\centerline{\psfig{file=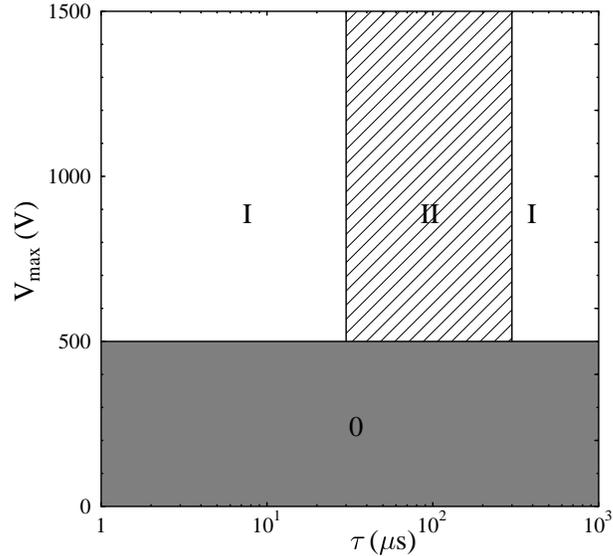,width=10.0cm,angle=0}}
\caption{``phase diagram'' of the Branly effect. The x-axis
  corresponds to the rise time of the voltage increase applied to the
  coheror. The y-axis corresponds to the maximum voltage used. In 0
  area ($V_{max} < 500  V$) no conducting path is created,
   in area I ($V_{max} > 500  V$ and $\tau < 30 \mu {\rm s\ or \ } \tau > 300$)
  only one single path is created and in area II ($V_{max} > 500  V$
  and $30 \mu s  \le \tau \le 300 \mu s$) the path(s) created
  can be either single or complex with loops, branches...}
\label{phas}
\end{figure}

The efficiency of this visualization technique being established, we
have then specifically studied the morphology of the conduction paths
induced by a DC voltage increase of varying rise time and maximum
voltage. The set-up consisted of a direct commutation limited by a
resistor-capacitor ($R_0C_0$) in series. The capacitance of the
coheror in its non-conducting state being measured to be $C
\simeq 0.5$ pF therefore we have used capacitances greater than C to
impose the rise time.  We have varied the maximum voltage from 300 V
up to 1500 V and the rise time $\tau=R_0C_0$ from 5 $\mu$s up to
about 100 ms. For each ($\tau$, $V_{max}$) point, we have performed 10
experiments.
\begin{figure}
\centerline{
\psfig{file=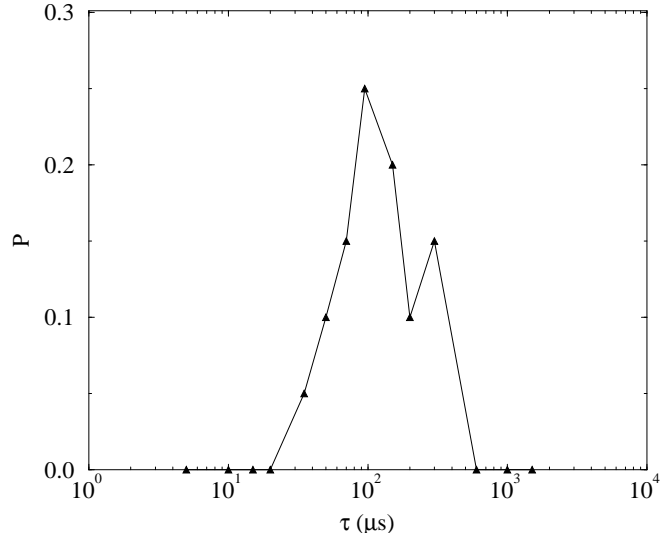,width=10.0cm,angle=-90}}
\caption{Probability $P$ of having complex paths versus the rise time $\tau$
  of the voltage increase applied to tg24
he coheror.}
\label{pbvst}
\end{figure}
The major results  are the following ones:\begin{itemize}
\item No path is created for maximum voltage values below a threshold
  $V_c \simeq 500\ V$. The coheror remains in its insulating state.
$V_c$ does apparently not depend on the rise time.

\item Both slow ($ \tau > 300\ \mu s$) and fast ($\tau < 30\ \mu
  s$) voltage rises lead to the creation of one simple unique
  conducting path (in a similar fashion as these shown on figure \ref{single}).

\item For intermediate rise time ($30\ \mu s \le \tau \le 300\ \mu s$,
  we generally see one simple conducting path but we frequently obtain
  (in a few tens of percents of the cases) complex paths with loops or
  branches and also non connected paths (see examples of such paths on
  figure \ref{mult}).
\end{itemize}
The probability of having complex conduction paths is plotted versus
the rise time of the voltage increase on figure \ref{pbvst}. This
figure has been obtained by averaging for each rise time the results of observations of
all experiments made with a maximum voltage higher than 500 V. The
behaviour of the coheror when applying a voltage increase is reported
in the ``phase diagram'' of the figure \ref{phas}.

\section{Discussion}

These preliminary results have obviously to be confirmed.  The
observation of two different ``phases'' is not completely
surprising. One could have expected however to keep complex paths for
very short rise times. Let us recall however that the system that we
have studied is quite small. It only holds 40$\times$40 grains, which
corresponds to a few pixels by grain with the 128$\times$128 we have
used. It would be naturally interesting to test larger systems.
Increasing the size of the system is however quite difficult in this
configuration. The separation between single and multiple paths can
also be difficult because of the presence of little loops. The latter
become undetectable when the resolution decreases.

The possibility of a visualization technique should allow new
developments in the study of this original phenomenon. One may first
use it for constructing a realistic model.  The main result of this
first work is the occurrence of multiple breakdowns. This point could be a very
interesting test for a future model. According to G. Kamarinos et al.,
the generation of paths is due to a succession of local dielectric
breakdowns of the oxide layers \cite{Kara75,Kama75,Kara90} (about 100 nm
thick in case of native aluminium). This theory suggests a
model for the medium as a simple network of resistor/capacitors.
Individual breakdowns would then occur when the local voltage reaches
a threshold level. If it can reproduce multiple breakdowns, such a
model should allow to relate the characteristic values of the
individual contacts with the critical rise time(s) for which multiple
breakdowns occur.

One can also think to further improvements of the visualization
technique. We have used here a DC voltage that allows us to see the
conducting path. If we use a sufficiently high frequency AC voltage, one
may hope to short-circuit some open contacts and also to visualize dead
ends (if some are present in the system).

This work has been devoted to the study of the insulator/conductor
transition of a metallic granular medium when the latter is submitted
to a DC voltage. As noticed by Branly \cite{Branly90}, such a
transition occurs not only when a high DC voltage is applied to the
system (this specific point was already described by 
Calzecchi-Onesti in 1884 \cite{Onesti}) but also at the reception of a
radio signal (that was the major contribution of Branly since it
permitted the realization of the first wireless telegraph
receivers). A natural extension of the 
visualization work that we have described in this paper is thus to
study this surprising property. On may especially try to compare the
breakdown patterns obtained in the two configurations allowing to
create a conducting path through the granular medium.

Other studies may  concern the influence of
mechanical stress on the conduction of a metallic granular medium.
But beyond further studies about the Branly effect, one can also think
of using infrared radiometry to visualize the mechanical stress
distribution in two-dimensional metallic granular medium
\cite{JPBouchaud96P} since the paths of maximum stress are also the
most~conductive.

\section*{Acknowledgements}
We acknowledge D. Fournier and E. Guyon for helpful discussions and
suggestions.

%\bibliography{../BIB/bibart,../BIB/phototherm}

%\end{multicols}
\end{document}